\documentclass[12pt]{article}

\input{tcilatex}

\begin{document}

\author{H. V. Fagundes and E. Gausmann \\
Instituto de F\'{i}sica Te\'{o}rica \\
Universidade Estadual Paulista \\
S\~{a}o Paulo, SP 01405-900, Brazil\\
e-mail: helio@ift.unesp.br, gausmann@ift.unesp.br}
\title{Cosmic Crystallography in Compact Hyperbolic Universes}
\date{23 November 1998}
\maketitle

\begin{abstract}
We try to apply the cosmic crystallography method of Lehoucq, Lachieze-Rey,
and Luminet to a universe model with closed spatial section of negative
curvature. But the sharp peaks predicted for Einstein-de Sitter closed
models do not appear in our hyperbolic example. So we turn to a variant of
that method, by subtracting from the distribution of distances between
images in the closed model, the similar distribution in Friedmann's open
model. The result is a plot with much oscillation in small scales, modulated
by a long wavelength quasi-sinusoidal pattern.
\end{abstract}

\section{Introduction}

The idea of cosmic crystallography was introduced by Lehoucq et al. \cite
{LeLaLu}, and applied by them to several Einstein-de Sitter (EdS) models
with closed (i. e., compact and boundless) spatial sections of Euclidean
geometry and nontrivial topology. The simplest of these cases has the
three-torus $T^{3}$ as spatial section - model E1 in Lachi\`{e}ze-Rey and
Luminet \cite{LaLu} review article on cosmic topology. This model provides
the sharpest example of the crystallographic effect of the topology: if $%
T^{3}$ is based on a cube of side $L$ as fundamental polyhedron (FP) then
the distances between images of a single source are in the form $%
d=(l^{2}+m^{2}+n^{2})^{1/2}L,$ with $l,m,n$ integers, so that a plot of the
distribution of distances $n(d)$ vs. $(d/L)^{2}$ provides neat peaks in the
integral values $l^{2}+m^{2}+n^{2}.$ In the other five EdS models with
closed spatial sections the peaks tend to be in smaller numbers or less
sharp, because the distances between some of the images of a single source
depend on the latter's position inside the FP. This is characteristic of
locally but not globally homogeneous models \cite{csc92}, and can be seen in
the study of model E4 by Fagundes and Gausmann \cite{FG98}: when $L=7200/h$
Mpc is of the order of magnitude of the observable universe's radius, no
peaks showed up in the distribution $n(d).$ These studies are based on
computer simulated catalogs, and cannot yet be compared with real catalogs,
given the present limitations of the latter.

The same absence of obvious peaks is true for a model with a closed
hyperbolic manifold $M^{3}$ as spatial section, and hence the metric of
Friedmann's open model. This is illustrated \ in Figs. 1 and 2, with $M^{3}$
the second of Best's \cite{best} three manifolds with an icosahedron as FP,
and listed as number $v2293(+3,2)$ in the census of closed, orientable
hyperbolic manifolds in Weeks's computer program SnapPea 2.5.3 \cite{snappea}%
. The figure shows plots of $n(d)$ against the distances in megaparsecs,
assuming the density parameter $\Omega _{0}=0.3$ and Hubble's constant $%
H_{0}=65$ km sec$^{-1}$Mpc$^{-1},$ and a radius of $12873$\ Mpc, equivalent
to maximum redshift $Z=1300,$ which we take to be the position of the
surface of last scattering (SLS) of the cosmic microwave background. Figures
1 and 2 show the plots for the distributions for the compact model and the
infinite model, respectively. Although there is more wiggling in the former,
the general shape is the same for both.

So we tried a further step in the crystallography method, which is to map
the \textit{differences }between the distributions for the compact and
infinite cases.

In Sec. 2 we explain our method and present the results, and in Sec. 3 the
latter are discussed.

\section{Calculations}

To fix our notation let the metric of the hyperbolic Friedmann model be
written as

\[
ds^{2}=dt^{2}-a^{2}(t)[d\chi ^{2}+\sinh ^{2}\chi (d\theta ^{2}+\sin
^{2}\theta d\varphi ^{2})]. 
\]

To simulate a random distribution of sources in the FP, we first looked for
such a distribution in the ball enclosed by the icosahedron's circumscribing
sphere, with radius $\chi _{out}=1.38257.$ This was done by dividing the
ball into 13 shells of thickness $\Delta \chi =\chi _{out}/13,$ with volume $%
\Delta V\approx 2\pi \sinh ^{2}\chi _{m}\Delta \chi ,$ where $\chi _{m}$ is
the medium radius of the shell; and also dividing the solid angle space into
ten zones, with $\Delta \Omega \approx 2\pi \sin \theta _{m}\Delta \theta $
in each zone, where $\Delta \theta =\pi /10$ and $\ \theta _{m}$ the zone's
medium value of $\theta .$ Then a number ($150$) of pseudorandom positions
in the ball were chosen by computer, with the condition that the occupations
in each shell and zone were proportional to $\Delta V$ and $\Delta \Omega $
respectively. Finally a program routine excluded from the obtained points
those that were outside the icosahedron. The $51$ remaining ones were
considered the positions of the sources in the FP. The radius of the SLS in
normalized units is $\chi =\chi _{\max }=2.33520$ for the assumed values of $%
\Omega _{0}$ and $H_{0}$.

In order to fill this space without gaps we needed $92$ copies of the
icosahedron, besides the FP itself. They were produced by the $20$
face-pairing generators $\gamma _{k}$ (one-letter words) of the covering
group $\Gamma $, $60$ two-letter, and $12\ $three-letter words in the$\
\gamma _{k}$'s. This set of icosahedra cover hyperbolic space up to a radius 
$\chi =2.33947$, slightly larger than $\chi _{\max }$. The $92$ operators
were then applied to each of the $51$ sources, the result being accepted as
an image if it lay inside the SLS ($\chi <\chi _{\max }$). The total of
potential images (they are potential in the sense that, because of evolution
and other factors, many of them may not be observable) thus obtained,
including the sources, was $1570$.

Then we looked for the present (or comoving) distances between each pair of
these images, by multiplying the normalized distances by the curvature
radius of the universe (see, for example, \cite{KT}), $R_{0}=cH_{0}^{-1}(1-%
\Omega _{0})^{-1/2}=5512.62$ Mpc. A list was prepared by grouping the
numbers of occurrence of distances in bins of $100$ Mpc width. The
corresponding plot is shown in Fig. 1.

We proceeded to obtain a similar distribution in Friedmann's open model,
with the same parameters as above, but with the same number of \
pseudorandom sources as there are images, i. e., $1570$ sources. A list
similar to the one above mentioned was plotted in Fig. 2. As indicated in
the Introduction, we could not see a significant difference between these
plots.

Finally we subtracted the values of the second list from those of the first.
Fig. 3 is a plot of these differences. One notices wild oscillations in the
scale of the bin width, modulated by a broad pattern, on the scale of $%
R_{0}. $ We fitted this plot with a five-term Fourier series of the form

\[
f(x)=\sum_{n=1}^{5}a_{n}\sin \frac{n\pi x}{\lambda }, 
\]
where $x=$ nearest integer to $d/(100$ Mpc), $\lambda =$ 258, and \{$a_{n}$%
\} = \{96.6, 134.4, -211.7, 127.0, -89.4\}.

\section{Discussion}

A \textit{Clifford translation} of a metric space $S$ is an isometry $g$ of $%
S$ such that \textit{distance}$(p,gp)$ is independent of point $p\in S$. Cf.
Wolf \cite{wolf}. In a recent paper, Gomero et al. \cite{gomero} link the
existence of sharp peaks in cosmic crystallography to that of elements of $%
\Gamma $ which are Clifford translations of the universal covering space of $%
M^{3}$. (See also Lehoucq et al. \cite{LeLuUz}.) But in Fig. 1 of \cite{FG98}
we see a significant peak that may not be produced by a Clifford
translation. The matter deserves further investigation.

In the case of the hyperbolic space $v2293(+3,2)$ treated here, there seems
indeed to be no sharp peaks, as predicted by \cite{gomero} and \cite{LeLuUz}%
. (It should be interesting to look for them in the case of a highly
symmetrical manifold, like the hyperbolic dodecahedron space \cite{ST}.)
This is why we proceeded to look for the differences in the distribution of
distances between this model and Friedmann's open model, whose plot does
show a significant pattern.

Of course we would not expect to be able to infer the correct cosmic
topology from a single plot, even one with sharp peaks. The topology of the
universe will probably be revealed slowly and progressively, as more and
better data are obtained, and several topology searching methods are applied
to them.

\medskip

E. G. thanks Conselho Nacional de Desenvolvimento Cient\'{i}fico e
Tecnol\'{o}gico (CNPq - Brazil) for a doctorate scholarship and financial
help for participation in this Symposium. H. V. F. thanks CNPq for partial
financial support, and Funda\c{c}\~{a}o de Amparo \`{a} Pesquisa do Estado
de S\~{a}o Paulo (FAPESP) for a grant to attend this event.

\bigskip

\bigskip \medskip \newpage

FIGURE CAPTIONS:

\bigskip

FIGURE 1. A plot of the number of occurences of the distances in the
abscissa, in bins of $100$ Mpc, for model with the finite, multiply
connected manifold $v2293(+3,2)$ as spatial section$,$ with a simulated
catalog of $51\,$sources and $1570$ potential images.

\bigskip

FIGURE 2. A plot similar to that of Fig. 1, now for the spatially infinite
Friedmann's open model and a simulated catalog of $1570$ sources
or\smallskip\ potential images.

\bigskip

FIGURE 3. The abscissa is the same as in Figs. 1 and 2, the ordinate is the
difference of their ordinates.

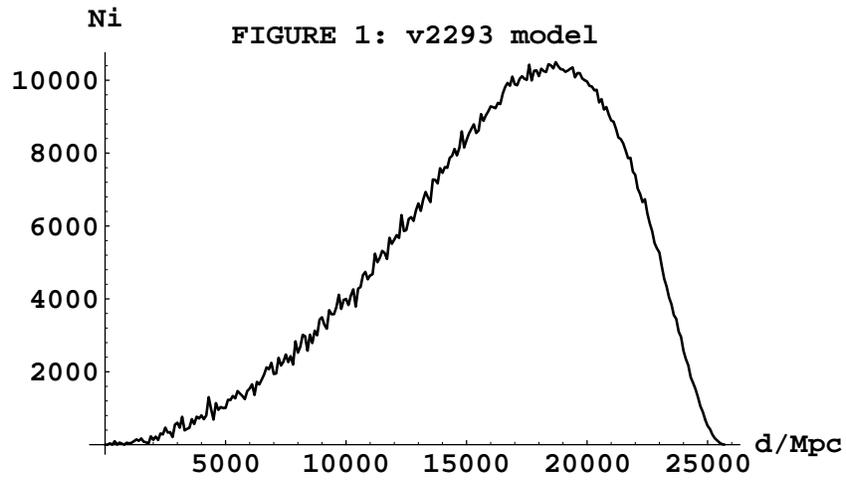
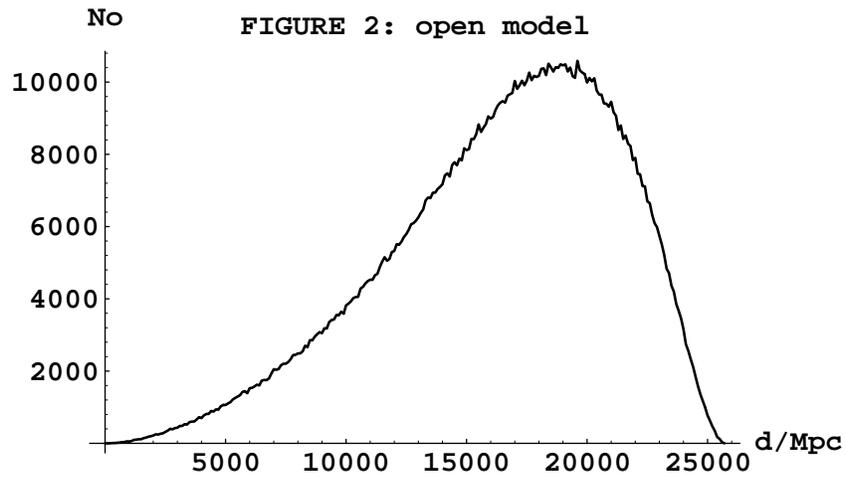
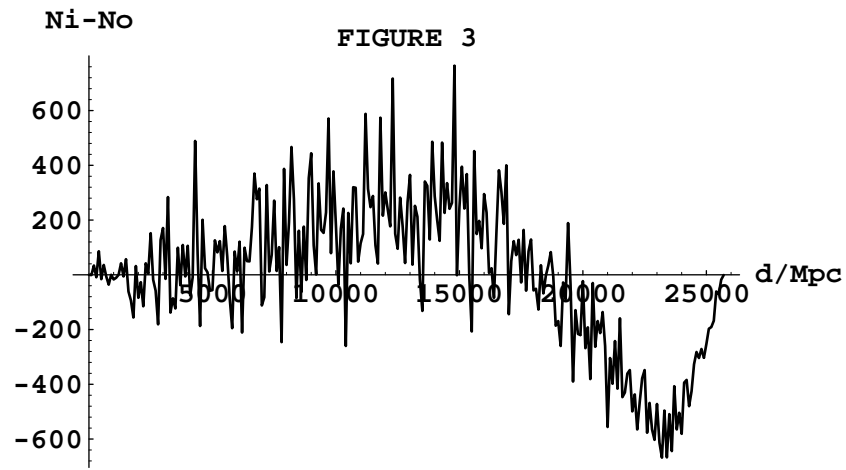
\end{document}